\documentclass{article}

\usepackage{amssymb,amsfonts,amsthm}
\usepackage{amsmath}
\usepackage{bm}
\usepackage[mathscr]{eucal}
\usepackage{wasysym}

\def\pmb#1{\setbox0=\hbox{$#1$}%
  \kern-.025em\copy0\kern-\wd0
  \kern.05em\copy0\kern-\wd0
  \kern-.025em\raise.0433em\box0}
\def\pmbs#1{\setbox0=\hbox{$\scriptstyle #1$}%
  \kern-.0175em\copy0\kern-\wd0
  \kern.035em\copy0\kern-\wd0
  \kern-.0175em\raise.0303em\box0}
\def\be{\begin{equation}}
\def\ee{\end{equation}}
\def\bea{\begin{eqnarray}}
\def\eea{\end{eqnarray}}

\def\hsp5{\hspace{5mm}}
\def\case#1/#2{\textstyle\frac{#1}{#2}}

\theoremstyle{plain}

\theoremstyle{remark}

\title{\bf \Large Composite spherically symmetric configurations in Jordan-Brans-Dicke theory\\}
\author{
\sc S.M.KOZYREV $^{}$\thanks{Electronic address: {\tt
sergey@tnpko.ru}}\ \\
$^{}${\small\em Scientific center gravity wave studies ''Dulkyn'',
PB 595, Kazan, 420111, Russia , Kazan, Russian Federation} }
\begin{document}
\maketitle
\begin{abstract}
In this article, a study of the scalar field shells in
relativistic spherically symmetric configurations has been
performed. We construct the composite solution of
Jordan-Brans-Dicke field equation by matching the conformal Brans
solutions  at each junction surfaces. This approach allows us to
associate rigorously with all solutions as a single glued "space",
which is a unique differentiable manifold $M^4$.
\end{abstract}

\section{Introduction}

    The field equations in relativistic field theories are non linear
in nature. For a classical field, the differential equations
consist of purely geometric requirements imposed by the idea that
space and time can be represented by a Riemannian (Lorentzian)
manifold, together with the description of the interaction of
matter and gravitation contained in Einstein's equations.

    This equation involved a match between a purely geometrical object
so called Einstein tensor G, and an object which depends on the
properties of matter fields the energy-momentum tensor $T$ which
contains quantities like the density and pressure. Hence, the
geometry of 4D spacetime is governed by the matter it contains.
However, this split is artificial.

The variables $ x$, used in Einstein's equations, represent
co-ordinates of points of abstract four-dimensional manifold Ì4.
In geometrical sense the different solutions of Einstein equations
generate different 4D pseudo-Riemannian space-time manifolds V4
(g). In other words, the emergent geometry is not a priori assumed
but defined from the solutions. Co-ordinates in each of such
spaces have the specific properties differing from their
properties in other spaces \cite{Gullstrand}.

    Einstein's equations determine the solution of
a given physical problem up to four arbitrary functions, i.e., up
to a choice of gauge transformations. Evidently, a structure of
spacetimes is mathematically represented by Einstein's equations
and four co-ordinate conditions \cite{Temchin}.
\begin{equation}
G_{i k }=T_{i k }, \label{eq1y}
\end{equation}
\begin{eqnarray}
C(\mu) g_{i k } &=&0. \label{eq2y}
\end{eqnarray}
where $g_{i k }$ metric tensor and $C (\mu )$ - some algebraic or
differential operators. Thereby for any four of components $g_{i k
}$ emerge the relations with remaining six and, probably, any
others, known functions. Certainly, equations (\ref{eq2y}) cannot
be covariant for the arbitrary transformations of independent
variables, and similarly should not contradict Einstein's
equations or to be their consequence. Moreover, in every class of
co-ordinates each time is postulated new subsystem (\ref{eq2y}).
Usually, four of ten field equations will not be transformed
according to any rules, but simply replaced by hand with the new.
Generally speaking, physics looks different in two different
classes of co-ordinates.

    It is well known, since the pioneering paper of Jordan \cite{Jordan}
 that the action of a scalar tensor theory is invariant under local transformations
 of units that are under general conformal transformations. This method of conformal
 transformation pointed out first by Pauli. The method provides a clear and powerful
 technique, free from mathematical ambiguity, but nevertheless requires careful
 consideration from the physical point of view. It should be stressed that the
 apparently different scalar-tensor theories can be mapped onto each other by
 conformal mappings \cite{Sokolowski}.

It is exactly the non-triviality of unambiguously
field-decomposition into a physical part, which represents the
true gravitational effect, and a pure geometric part, which
represents the feigned gravitational effect coupled with choice
classes of co-ordinates. Moreover, scalar tensor theories are
mathematically equivalent to general relativity and there are
convincing arguments that they are also physically equivalent to
it \cite{Magnano}, \cite{Sokolowski}. In other terms the
conformally invariant scalar may be represented in disguise in
infinite number of ways as commonly viewed as special kinds of
matter. Clearly, there is no trace of concerning what conformal
frame we live in, and on what physical grounds we are able to
select which.

\section{Field equations.}

A general Lagrangian for a scalar-tensor gravity theory,
\begin{eqnarray}
 L =\sqrt{g} \left[f(\phi) R -\frac{\omega (\phi) }{\phi}g^{ik}\phi_{,i}
 \phi_{,k}+ 2\phi V(\phi) \right] + L_m \label{eq1f}
\end{eqnarray}
where $f(\phi) > 0$ and $L_m$ is the Lagrangian density describing
"ordinary" matter as opposed to the scalar field $\phi$, which
effectively plays the role of a form of non-conventional matter in
the field equations. Here $g_{ik}$ is the metric tensor with
determinant g, $f(\phi)$  and $ \omega (\phi) $ are arbitrary
coupling functions, $\phi$  is the scalar field with potential $
V(\phi)$. The Lagrangian (\ref{eq1f}) is invariant under the
conformal transformations
\begin{eqnarray}
 g_{i k }^* =\phi ^{2 \zeta} g_{i k }, &  \zeta \ne  1,  & \phi ^* = \phi ^{1-\zeta} \label{eq3y}
\end{eqnarray}
this is equivalent to the transformation applied to a line
element,
\begin{eqnarray}
 d s^{* 2} =\phi ^{2 \zeta}  d s^2.  \label{eq4y}
\end{eqnarray}
The case $\zeta =  1$, represents just a self-interacting scalar
$\phi$ minimally coupled to gravity and is designated as Einstein
conformal frame, with
\begin{eqnarray}
g^*_{ i k } =e^{2 \zeta \phi} g_{i k }, \phi^* =\kappa e^{2 \zeta
\phi} \label{eq2f}
\end{eqnarray}
In the literature the Jordan frame (\ref{eq1f}) and the Einstein
frame (\ref{eq2f}) are those discussed most frequently.

The Jordan frame in which the theory is formulated is the set of
dynamical variables $g_{ik}, \phi$ describing the gravitational
field Lagrangian can then take, among others, the equivalent forms
\begin{eqnarray}
 L =\sqrt{g} \left[f(\phi^*) R^* +  V^*(\phi^*) \right]
 =\sqrt{\stackrel{-}{g}}
 \left[\stackrel{-}{R}(\stackrel{-}{g}) -\stackrel{-}{g}^{ik}\stackrel{-}{\phi_{,i}}
 \stackrel{-}{\phi_{,k}}-\stackrel{-}{V}(\stackrel{-}{\phi}) \right] \label{eq3f}
\end{eqnarray}

 Thus the theory can be expressed
 in terms of infinite number of conformally related frames; in such a system one can
 always make a field redefinition (a change of variables) to another conformal frame
 via a conformal mapping. Here we restrict out attention to Jordan-Brans-Dicke (JBD) theory,
where $\omega (\phi)= const, V(\phi)=0$. From the above action we
can find the JBD field equations
\begin{equation}
R_{i k }-\frac 12Rg_{i k }=\frac 1 {\phi \ }T_{i k}^m+ T_{i k
}^{JBD},  \label{eq4f}
\end{equation}
where

\begin{eqnarray}
T_{i k }^{JBD} &=& [ \frac \omega {\phi ^2}\left( \nabla _i \phi
\nabla _k \phi -\frac 12g_{i k }\nabla _j \phi
\nabla ^j \phi \right) +  \nonumber \\[0.01in]
&&\ \ +\frac 1\phi \left( \nabla _i \nabla _k \phi -g_{i k }\nabla
_j \nabla ^j \phi \right) ].  \label{eq5f}
\end{eqnarray}
and

\begin{equation}
\nabla _j \nabla ^j \phi =\frac{T_k ^{m\ k }}{3 + 2 \omega },
\label{eq6f}
\end{equation}
where $T^{JBD}_{ik} $ is often identified with an effective
stress-energy tensor of the scalar field $\phi$. The metric tensor
satisfies to those or other co-ordinate conditions if some of
quantities $g_{i k }$ are linked by some relations, - whether it
be in any point, on a surface or in four-dimensional domain
$\Omega \subset$ $V^4$(g). By definition all co-ordinate systems
in manifold $M^4$ at least locally are equivalent; on the other
hand if in $M^4$ the metric is introduced, properties of functions
$g_{i k }$ in different co-ordinates become different.

In relativistic scalar tensor theory, a covariant presentation of
the matching conditions, across a separating hypersurface,
requires the continuity of the first and second fundamental forms.
Let us consider two distinct manifolds $M^4$$^+$ and $M^4$$^-$.
The metric in these manifolds generated by set of solutions of
field equations (\ref{eq4f}) given by $g_{i k }^+$(
\textit{x}$^i$$_+$) and $g_{i k }^-$( \textit{x}$^i$$_-$), in
terms of independently defined coordinate systems
\textit{x}$^i$$_+$ and \textit{x}$^i$$_-$. The manifolds glued at
the boundary hypersurfaces $\Sigma$ $_+$ and $\Sigma$ $_-$  using
independently defining co-ordinates systems \textit{x}$^i$$_\pm$.
A common manifold $M = M^+ \bigcup M^-$ is obtained by assuming
the continuity of four-dimensional coordinates
\textit{x}$^i$$_\pm$ across $\Sigma$, then $g_{i k }^+$ = g$_{i k
}^-$ is required, which together with the continuous derivatives
of the metric components $\partial$ $g_{i k }$
/$\partial$\textit{x}$^j$ $\mid_+$ = $\partial$ $g_{i k }$
/$\partial$\textit{x}$^j$ $\mid _-$, provide the Lichnerowicz
conditions \cite{Lichnerowicz}.

The resulting manifold $M$ is geodesically complete and possesses
two regions connected by a hypersurface $\Sigma$. The extrinsic
curvature, or the second fundamental form, is defined as

\begin{eqnarray}
K^i_{k \pm} =\frac{1}{2}g^{i k}\frac{\partial g_{k i}}{\partial
\eta} \mid _{\eta=\pm 0} \nonumber
\end{eqnarray}
where $\eta$ the proper distance away from the $\Sigma$.

\section{Composite solution in Jordan-Brans-Dicke theory.}

In the present article we will assume the static spherically
symmetric configurations involved the matching of conformally
invariant interior "vacuum" solutions. The solution will be given
in terms of explicit closed-form functions of the radial
coordinate for the three metric coefficients. The physical and
geometrical meaning of the coordinate $ r $ is not defined by the
spherical symmetry of the problem and is unknown a priori
\cite{Fiziev}. Note, that its choice has been discussed from
physical point of view by Eddington as early as in
\cite{Eddington}. According to the widespread common opinion, the
most common form of line element of a spherically symmetric
spacetime in comoving coordinates can be written as
\begin{eqnarray}
ds^2=-g_{t t } (r, t) dt^2+  g_{r r }(r, t) dr^2 + 2g_{r t }(r,
t)dr dt + \rho(r, t) ^2 (d\theta^2 + \sin^2(\theta) d \varphi^2).
\label{eq5y}
\end{eqnarray}
We are free to reset our clocks by defining a new time coordinate
\begin{eqnarray}
t=t'+ u(r). \nonumber
\end{eqnarray}
with \textit{u(r)} an arbitrary function of \textit{r}. This
allows us to eliminate the off-diagonal element g$_{r t }$.
Therefore we shall consider the matching of two static and
spherically symmetric spacetimes given by the following line
elements
\begin{eqnarray}
ds_{\pm}^2=-g_{t t } (r, t)_{\pm} dt^2+  g_{r r }(r, t)_{\pm} dr^2
+ \rho(r, t)_{\pm} ^2 (d\theta^2 + \sin^2(\theta) d \varphi^2).
\label{eq6y}
\end{eqnarray}
of $M^4 _\pm$, respectively, where $g_{t t }(r, t)$, $g_{r r }(r,
t)$ and $\rho(r, t)$  are of class C$^2$.

In the static spherically symmetric case the choice of spherical
coordinates and static metric dictates the form of three of the
gauge fixing coefficients (\ref{eq6y}):
\begin{eqnarray}
\Gamma_t =0, \Gamma_\theta = -cot \theta, \Gamma_\varphi = 0,
\end{eqnarray}
where
\begin{eqnarray}
\Gamma_i =-\frac{1}{\sqrt{g} }g_{i k } \partial_j (\sqrt{g}g^{j k
}) ,
\end{eqnarray}
but the form of the $\rho(r)$ are still not fixed. In the
literature one can find different choices of the function
$\rho(r)$  but the isotropic class of coordinates is those
discussed most frequently. The static spherically symmetric matter
free solution of  JBD theory in isotropic coordinates is given by
\cite{Brans}.

It is well known, that among the four classes of the Brans static
spherically symmetric solution of the vacuum JBD theory of gravity
only two are really independent \cite{Bhadra}. It should be noted
that the JBD action has a conformal invariance (\ref{eq3y})
characterized by a constant gauge parameter $\zeta $. Arbitrary
value of  $\zeta $ can actually lead to shift from the value of
Brans solutions and change scalar field and metric coefficients.
The general solutions for static spherically symmetric
configurations in isotropic class of coordinates
\begin{eqnarray}
ds^2=-e^{2 \alpha (r)} dt^2+  e^{2 \beta (r)} [dr^2 + \rho(r) ^2
(d\theta^2 + \sin^2(\theta) d \varphi^2)]. \label{eq7y}
\end{eqnarray}
then looks like \cite{Bhadra2}:

\begin{eqnarray}
e^{\alpha (r)} =
e^{\alpha_0}\left(\frac{1-\frac{B}{r}}{1+\frac{B}{r}
}\right)^{\frac{1+\zeta C}{\lambda}},\nonumber \\
e^{\beta (r)}=e^{\beta_0} \left({1+\frac{B}{r} } \right)^2
\left(\frac{1-\frac{B}{r}}{1+\frac{B}{r}
}\right)^{\frac{\lambda-C-1+\zeta C}{\lambda}}\label{eq8y} \\
 \phi(r) = \phi_0 \left(\frac{1-\frac{B}{r}}{1+\frac{B}{r}
}\right)^{\frac{C(1-2 \zeta )}{\lambda}},\nonumber
\end{eqnarray}

and
\begin{eqnarray}
e^{\alpha (r)} = e^{\alpha_0} e^{\frac{2(1+\zeta C)}{\lambda}arctan(\frac{r}{B})},\nonumber \\
e^{\beta (r)}=e^{\beta_0} \left( 1+\frac{B^2}  {r^2}\right) e^{\frac{2(C+1-\zeta C)}{\lambda}arctan(\frac{r}{B})}\label{eq10y} \\
 \phi(r) = \phi_0 e^{\frac{2C(1-2 \zeta
)}{\lambda}arctan(\frac{r}{B})},\nonumber
\end{eqnarray}
where the constants $\lambda, C, \zeta $ are still connected via
the JBD field equations by
\begin{eqnarray}
\lambda ^2=1+C+\frac{1}{2}C^2\left[ 2+\omega -2 \zeta (3+2\omega)+
\zeta ^2 (6+4 \omega) ) \right]. \label{eq9y}
\end{eqnarray}
For our advance to the composite spherical configurations, we look
at the families of scalar field shells. The scalar layers
generated in this approach are obtained by matching of scalar
field and introducing a discontinuity in its derivative. In this
case the boundary surface entails via the field equations a jump
in second derivations of metric coefficient, but metric
coefficient and its first derivatives remains continuous. Now in
order to justify calling the geometry a composite configuration we
need an explicit definition for the constants in the solutions
(\ref{eq8y})-(\ref{eq10y}). At the each boundary surface we have
five equations for five unknown variables $B, C, \alpha_0,
\beta_0, \phi_0 $. Thus, from the junction conditions, the each
interior metric and scalar field parameters can be determined at
the boundary surface in terms of the respective exterior metric
and scalar field parameters $ \zeta$. Consequently, once the
constants of the ambient space are fixed, the property in the
interiors layers region roughly corresponds to the values inferred
from observations. The brief computation of yields:
\begin{eqnarray}
B_{int}=\pm \sqrt {\frac{r_s^3[X+Y+Y C_{int}(X-Y)\zeta_{int}
C_{int}]}{2+2C_{int} \zeta_{int} +r_s[X+Y+Y
C_{int}(X-Y)\zeta_{int} C_{int}] }}, \label{eq11y}
\end{eqnarray}
where
\begin{eqnarray}
&&X=\frac{2B_{ext} (-B_{ext} \lambda_{ext} + r_s(1+C_{ext}-C_{ext}
\zeta_{ext}))}{r_s (B_{ext}^2-r_s^2)\lambda_{ext}} , \nonumber\\
&&Y=-\frac{2B_{ext} (1+C_{ext} \zeta_{ext})}{
(B_{ext}^2-r_s^2)\lambda_{ext}}. \label{eq12y}
\end{eqnarray}
All terms in (\ref{eq11y}), (\ref{eq12y}) with index $int$
contribute to the internal and $ext$ external scalar layers, $r_s$
denotes radius of boundary hypersurface.

 In this case we deal with non-linear equations for the
unknown variables; the situation with finding constant of
integration $C_{int}$ is more complicated. It seems that numerical
integration is the "simplest" way to treat this problem. To this
end, following this approach we regard exterior region of JBD
composite spherically symmetric configurations by the
Schwarzschild metric or a flat space. As a particular but
interesting example, we consider the case where gauge parameter
$\zeta =1/C$. This value of parameter   make possible to find a
field configurations in which the inside of the shell is conformal
Brans I or Brans II space while the outside is a flat space. It
easy to show that the arbitrary constants are determined by using
matching conditions and the constant $B$ determined for conformal
Brans I as follows:
\begin{eqnarray}
B=r_s \sqrt {\frac{2}{ 2+\omega}}, \label{eq13y}
\end{eqnarray}
and for conformal Brans II
\begin{eqnarray}
B=-r_s \sqrt {\frac{2}{ -2-\omega}}, \label{eq13y}
\end{eqnarray}

 Moreover, we can tune the scalar field of the outside of the
shell; while of the inside will be flat space. This solution is
asymptotically flat and hence that connects two flat regions of
space. An interesting result in this case is that the flat space
produces Keplerian mass of the scalar configurations. Otherwise,
for  $\zeta =1/C$ we might treat each of solutions (\ref{eq8y}),
(\ref{eq10y}) as independently derived solution matched to a flat
space. One now has three connected regions, that is, one-side
conformal Brans I, a both-side flat region and another one-side
Brans I with different arbitrary constant or Brans II region.

We stress the fact that for all above examples the Darmois-Israel
junction conditions \cite{Israel} are fulfilled.

\section{A solutions with throat.}

The metric coefficients in conformal Brans solutions are required
to satisfy some constraints, enumerated in \cite{Visser}, in order
that they have a throat. The important point is that the metric
coefficients in Eqs. (\ref{eq8y}) and (\ref{eq10y}) depend on
three parameters $ \omega , \zeta, C $ and satisfying the
inequality
\begin{eqnarray}
1+C+\frac{1}{2}C^2\left[ 2+\omega -2 \zeta (3+2\omega)+ \zeta ^2
(6+4 \omega) ) \right]>0. \label{eq1t}
\end{eqnarray}
 Confronting the isotropic Brans I
solution with the conformal Brans I metric, the parameter $
\lambda $ is defined as
\begin{eqnarray}
\lambda ^2=1+C+\frac{1}{2}C^2\left( 2+\omega  \right),
\label{eq2t}
\end{eqnarray}
so that in the case of conformal solutions we have additional
degree of freedom. This implies that the range of $\omega$ is
dictated by the range of $ C $ and $\zeta$ which, in turn, is to
be dictated by the requirements of throat geometry. In this
context, the violation of the weak energy condition combined with
an adequate choice of C and   could provide a viability of
"wormhole" spacetime and less restrictive interval for $\omega$
from the case of $ \frac 32 < \omega < \frac 43 $ considered in
\cite{Nandi}.

The new results include matching between solutions with throat and
other conformal Brans and after that Schwarzschild solutions.
First of all we point out that in this model the stars acquires
features of a many-component objects (shells of "scalar gas" with
different properties) whose distribution in the observed
3-dimensional volume can has. Moreover, such a picture can
represents a Schwarzschild background, while the interior should
be considered as vacuum solution of JBD which defined a Keplerian
mass of this object.

  In scheme presented in this report, studies of possible "wormhole" solutions in alternative
gravitation was thought of as a way of understanding the role of
different fields as the "carrier" of exoticity together with the
aim of finding phenomena for which different qualitative behaviors
to those of standard General Relativity model may arise.

\section{Summary}

The contents of the paper may be summarized as under:

 (1) The conformal invariance of the vacuum JBD theory shows that the
solutions in it are not unique. A conformal transformation between
Brans solutions can be interpreted as a change of local units of
length \cite{Brans}. On the other hand, scalar fields are commonly
viewed
 as an abnormal kind of matter originates exactly from this energy-momentum
 tensor. One can in principle assume gauge-dependence of right-hand-side of
 equation (\ref{eq5f}) as a variety of matter fields with different equations of state.
 Due to gauge invariance of JBD field equations, it is possible to generate
 and infinite set of solutions simply by assigning arbitrary values to the
 gauge parameter $\zeta$ and the resulting sets of solutions form families of
 spacetimes having the same physical content.

(2) As we have seen, even restricting to static spherically
symmetry JBD theory has number of conformal solutions. Then one
consequence of this is the possibility to use these conformal
solutions as reservoirs shells of "scalar gas" of a "star" match
it each other and finally with Schwarzschild metric or a flat
spacetime.

(3) The JBD theory contains some regions of the parameter space in
which the scalar field may play the role of exotic matter,
implying that it might be possible to build a wormholelike
spacetime with the presence of "scalar gas" at the throat. Varying
the real gauge parameter $\zeta $, one can obtain value of
$\lambda $ from any given $\omega $ on either side of the divide
$\omega = -3/2$ but not across it since it is the fixed point of
the relation (\ref{eq1t}).

  It should also be noted that because Jordan-Brans-Dicke is a highly
non-linear theory, it is not always easy to understand what
qualitative features solutions might possess, and here the
composite class of solutions can used an a guide.


\begin{thebibliography}{99}
\bibitem{Gullstrand}  {\small A. Gullstrand, Allgemeine Losung des statischen Eink?rperproblem in der Einsteinschen Gravitationstheorie. Ark. for Mat., Astr. o. fysik, Bd.16, N 8,  (1921) }

\bibitem{Temchin}  {\small A.N.Temchin, Uravneniia Einshteina Na Mnogoobrazii, Moskow, URSS,1999, (Russian). },

\bibitem{Jordan}  {\small P. Jordan, Schwerkraft und Weltall, {\bf (Braunschweig) },
(1955).  }

\bibitem{Magnano}  {\small G. Magnano, L.M. Sokolowski, Physical equivalence between nonlinear gravity theories and a general-relativistic self-gravitating scalar field,  Phys. Rev. D 50, 5039 (1994).

}

\bibitem{Sokolowski}  {\small L. M. Sokolowski, Universality of Einstein's general Relativity, Talk given at the 14th conference on general relativity and gravitation Florence (Italy), arxiv:gr-qc/9511073, (1995).}

\bibitem{Lichnerowicz}  {\small A. Lichnerowicz, "Theories Relativistes de la Gravitation et de l'Electromagnetisme," Masson, Paris (1955).}

\bibitem{Fiziev}  {\small P. Fiziev, Gravitational Field of Massive Point Particle in General Relativity, gr-qc/0306088, ICTP preprint IC/2003/122. P.P. Fiziev, T.L. Bojadjiev, D.A. Georgieva, Novel Properties of Bound States of Klein-Gordon Equation in Gravitational Field ofMassive Point, gr-qc/0406036. P. Fiziev, S. Dimitrov, Point Electric Charge in General Relativity, hep-th/0406077. P. Fiziev, On the Solutions of Einstein Equations with Massive Point Source, gr-qc/0407088. }

\bibitem{Eddington}  {\small A. S. Eddington, The mathematical theory of relativity, 2nd ed. Cambridge, University Press, 1930 (repr.1963).)}

\bibitem{Brans}  {\small C. Brans and R. H. Dicke, Phys. Rev. {\bf 124 }, 925-935
(1961).}

\bibitem{Bhadra}  {\small A. Bhadra and K. K. Nandi,  Mod. Phys. Letts. 16, 2079 (2001). }

\bibitem{Bhadra2}  {\small A. Bhadra, I. Simaciu, K. K. Nandi, Y.-Z. Zhang,  Comments on new Brans-Dicke wormholes, Preprint arXiv:gr-qc/0406014, (2004)}

\bibitem{Israel}  {\small W. Israel, "Singular hypersurfaces and thin shells in general relativity," Nuovo Cimento 44B, 1 (1966); and corrections in ibid. 48B, 463 (1966).}

\bibitem{Visser}  {\small M. Visser, Phys. Rev. D 39, 3182 (1989); Nucl. Phys. B328, 203 (1989); Lorentzian Wormholes-From Einstein To Hawking (AIP, New York, 1995).}

\bibitem{Nandi}  {\small K. K. Nandi, B. Bhattacharjee, S. M. K. Alam and J. Evans, Phys. Rev. D 57, 823 (1998).}




\end{thebibliography}
\end{document}